\documentclass{aa}  

\pdfoutput=1 
\usepackage{graphicx}
\usepackage{dcolumn}
\usepackage{bm}
\usepackage{xspace}
\usepackage{xcolor}
\usepackage{hyperref}
\usepackage[mathlines]{lineno}
\usepackage{physics}
\usepackage{booktabs}
\usepackage[normalem]{ulem}

\newcommand{\AREPO}{\textsc{arepo}\xspace}

\newcommand{\msol}{\,$M_\odot$\xspace}

\newcommand{\rsol}{\,$R_\odot$\xspace}
\newcommand{\E}[1]{\times10^{#1}\xspace}

\makeatletter
\renewcommand*\aa@pageof{, page \thepage{} of \pageref*{LastPage}}
\makeatother

\begin{document}


\title{Gravitational wave emission from dynamical stellar interactions}
\author{
    Javier Mor\'an-Fraile\inst{1,}\thanks{\email{javier.moranfraile@h-its.org}}
    \and
    Fabian R. N. Schneider\inst{1,2}
    \and
    Friedrich K. Röpke\inst{1,3}
    \and
    Sebastian T. Ohlmann\inst{4}
    \and
    Rüdiger Pakmor\inst{8}
    \and
    Theodoros Soultanis\inst{5,1,6}
    \and
    Andreas Bauswein\inst{5,7}
    }

\institute{
    Heidelberger Institut für Theoretische Studien (HITS),
              Schloss-Wolfsbrunnenweg 35, 69118 Heidelberg, Germany
    \and
    Zentrum f{\"u}r Astronomie der Universit{\"a}t Heidelberg,
    Astronomisches Rechen-Institut,
    M{\"o}nchhofstr.\ 12-14, 
    69120 Heidelberg, Germany
    \and
    Zentrum f\"ur Astronomie der Universit\"at Heidelberg,
    Institut f\"ur Theoretische Astrophysik, 
    Philosophenweg 12,
    69120 Heidelberg, Germany
    \and
    Max Planck Computing and Data Facility, 
    Gie{\ss}enbachstra{\ss}e 2, 
    85748 Garching, Germany
    \and
    GSI  Helmholtzzentrum  f\"ur  Schwerionenforschung,  Planckstra{\ss}e  1,  64291  Darmstadt,  Germany
    \and
    Max-Planck-Institut f\"ur Astronomie, K\"onigstuhl 17, 69117 Heidelberg, Germany
    \and 
    Helmholtz Research Academy Hesse for FAIR (HFHF), GSI Helmholtz Center for Heavy Ion Research, Campus Darmstadt,  Germany
    \and
    Max-Planck-Institut f\"ur Astrophysik,
    Karl-Schwarzschild-Str. 1, 85748 Garching,
    Germany
    }

\date{\today}

\abstract{
We are witnessing the dawn  of gravitational wave (GW) astronomy. With currently available detectors, observations are restricted to GW frequencies in the range between ${\sim} 10\,\mathrm{Hz}$ and $10\,\mathrm{kHz}$, which covers the signals from mergers of compact objects. The launch of the space observatory LISA will open up a new frequency band for the detection of stellar interactions at lower frequencies. In this work, we predict the shape and strength of the GW signals associated with common-envelope interaction and merger events in binary stars, and we discuss their detectability. Previous studies estimated these characteristics based on semi-analytical models. In contrast, we used detailed three-dimensional magnetohydrodynamic simulations to compute the GW signals. We show that for the studied models, the dynamical phase of common-envelope events and mergers between main-sequence stars lies outside of the detectability band of the LISA mission. We find, however, that the final stages of common-envelope interactions leading to mergers of the stellar cores fall into the frequency band in which the sensitivity of LISA peaks, making them promising candidates for detection. These detections can constrain the enigmatic common-envelope dynamics. Furthermore, future decihertz observatories such as DECIGO or BBO would also be able to observe this final stage and the post-merger signal, through which we might be able to detect the formation of Thorne-\.Zytkow objects.}

\maketitle

\section{\label{sec:intro}Introduction}
Mass transfer between stars in binary systems can lead to many different outcomes. Some of these outcomes can dramatically change the systems \citep{Langer2012a,DeMarco2017}. Whenever the stars that form a binary get close enough to each other so that one of them fills its Roche lobe (because of either a change in the evolution stage of the stars or in the orbital parameters), mass transfer will start. Unstable mass transfer can become dynamical and lead to a common-envelope (CE) event or even a full stellar merger. In a CE event, the accretor is engulfed in the envelope of the giant donor \citep{Paczynski1976} and the stellar cores transfer orbital energy to the envelope. This shrinks the orbital separation. The energy injected in the envelope can lead to its ejection (successful CE event). As a result, the orbit will cease shrinking, and the outcome of the interaction is a binary system in a tighter orbit \citep{Ivanova2013}. In contrast, if the companion fails to eject the envelope of the giant star, it will continue to inspiral and ultimately result in a merger of the core of the giant star and the companion (CE merger).

With the upcoming launch of the Laser Interferometer Space Antenna (LISA) \citep{Baker2019}, we may have a new tool for the study of these dynamical phases of CE events and stellar mergers, as they are promising sources for gravitational wave (GW) signals \citep{Ginat2020, Renzo2021}. Compared to the detectors of the LIGO\footnote{Laser Interferometer Gravitational-Wave Observatory}-Virgo-KAGRA\footnote{Kamioka Gravitational Wave Detector} network \citep{Aasi+2015,Acernese+2014,Akutsu2021}, which operates in a frequency band centered on hundreds of Hz, LISA will have a peak sensitivity centered at $\sim 4 \mathrm{mHz}$. The gap in between those frequency bands is proposed to be covered in the future with decihertz observatories such as the DECi-hertz Interferometer Gravitational wave Observatory (DECIGO) \citep{Seto2001} or the Big Bang Observer (BBO) \citep{Phinney2003}.
The frequency range of LISA will allow it to detect signals from wider binary systems (the frequency of the GWs released by a binary system is twice the orbital frequency: $f_{\mathrm{GW}} = 2\times f_{\mathrm{orb}}$). Similarly to the case of compact object mergers, the GWs produced by a stellar binary system shift to higher frequencies as the orbit shrinks.
In the case of stellar mergers and CE events, the rate at which the orbital separation decreases is generally not governed by the emission of GW radiation, but is mainly driven by mass transfer and drag forces.
Although not dynamically dominant, GWs provide an interesting opportunity for observing such short-lived and elusive events.

Three-dimensional hydrodynamic (3D HD) or magnetohydrodynamic (3D MHD) simulations are one of the most reliable (albeit expensive) ways of modeling CE events \citep[see][for some recent examples]{Ricker2012, Ohlmann2016, Staff2016, Prust2019, Reichardt2019, Sand2020, Chamandy2020, Reichardt2020, Moreno2021, Glanz2021a, Lau2022, Lau2022a, Zou2022, Ondratschek2022} and mergers between different types of stars, such as main-sequence (MS) stars \citep{Schneider2019} and white dwarfs \citep[][]{Munson2021,Dan2011,Zhu2015}. 
In setups in which general relativity (GR) is dynamically not relevant, these methods provide the most consistent predictions of the merger process, and GWs can be derived from these simulations, but this has rarely been done.
In the case of white dwarf mergers, predictions of the GWs that are released were obtained using smoothed particle hydrodynamics \citep[][]{Rasio1994,Loren-Aguilar2005}, as they are a known target for LISA \citep{Baker2019}. Most of the research done has focused on the GWs that are emitted during the inspiral phase rather than during the merger, however.
There are predictions of the GWs emitted during CE events \citep{Ginat2020, Renzo2021}, but because it is difficult to fully model them, all of the research so far has been based on analytic or semi-analytical models. 
The reliability of these predictions is limited by the fact that the interaction between the inspiraling companion and the envelope is not modeled in a self-consistent 3D MHD approach. Consequently, they ignore the contribution of the envelope itself to the production of GWs and employ an approximate drag force that drives the inspiral.

For neutron-star mergers, however, 3D HD simulations accounting for GR effects are common \citep[][]{Baiotti2017,Bauswein2019,Dietrich2021,Shibata2015}. 
This has broadly been applied for the study of GW signals that are produced by neutron-star mergers as they can be detected by the LIGO-Virgo-KAGRA network.

In this work, we fill the gap and derive GW signals from up to date 3D MHD simulations of binary interactions for which GR is not dynamically relevant.
We revisit an already published simulation of a main-sequence star merger \citep{Schneider2019} and run two new simulations of a successful CE ejection and a CE merger.

In Sect.~\ref{sec:methods} we briefly describe the methods we used for the different simulations and how we computed the GW signals.
In Sect.~\ref{sec:results} we explain the outcome of the simulations and discuss the GW signals.
We discuss the detectability of these events in Sect.~\ref{sec:discussion}, and we summarize our main findings in Sect.~\ref{sec:conclusion}.

\section{Methods\label{sec:methods}}
All simulations were carried out with the code \AREPO \citep{Springel2010,Pakmor2011}, which is a 3D MHD code that uses an unstructured moving Voronoi mesh with a second-order finite-volume approach. Gravity is Newtonian and was computed using a tree-based algorithm. The energy loss by GWs is negligible and was therefore not taken into account. 
For the successful CE ejection simulation, we included the impact of recombination energy by using the OPAL \citep{Rogers1996,Rogers2002} equation of state (EoS). We also used the OPAL EoS for the main-sequence star merger, whereas for the CE merger, in which we consider only the stellar cores, for which ionization effects are irrelevant, we used the Helmholtz EoS \citep{Timmes2000}.
The initial models of the main-sequence stars as well as the giant star were created using the one-dimensional stellar evolution code MESA \citep{Paxton2011, Paxton2013, Paxton2015}.
The setups of the different simulations are summarised in Table \ref{tab:setups}.

\subsection{Setups\label{sec:setups}}

\subsubsection{Main-sequence star merger}
For the merger between main-sequence stars, we used the simulation described in \cite{Schneider2019}, involving two early main-sequence stars with masses of 8\msol and 9\msol with radii of 1.7\rsol and 1.9\rsol, respectively.  It is computationally not feasible to simulate the merger from the point of Roche-lobe overflow until the merger, so the merging process was sped up artificially for 1.5 orbits, and only then was the evolution of the merger followed by the simulation.

\subsubsection{Successful CE ejection}
In giant stars, the dynamical timescale of the outer parts of the envelope is orders of magnitude longer than in the core. Accurately resolving the core requires that the simulation uses time steps on the order of milliseconds, whereas dynamical processes involving the envelope will occur on a timescale of days.
For this reason, the computational cost of simulating the common-envelope phase with a fully resolved core and companion becomes too expensive, and therefore, approximations are necessary.

For the successful CE ejection, we used one of the models described in \cite{Ohlmann2016thesis} as primary star. We took a model with a zero-age main-sequence (ZAMS) mass of 2\msol, evolved with MESA until the red giant (RG) stage, at which point it had a radius of 49\rsol. This model was mapped onto an \AREPO grid. For this simulation, we used a version of \AREPO with the same methods and the OPAL EoS as in \cite{Sand2020}. We replaced the core of the giant star and the companion with a gravitation-only particle to limit the range in timescales, resolving  their softening lengths with a minimum of 20 cells following \cite{Ohlmann2016thesis}. The softening length of both particles was initially set to $2.47$\rsol, and this value was reduced during the simulation to ensure that it remained below 40\% of the separation between the core and the companion. We did not include magnetic fields in this simulation. The masses of the core and companion were 0.37\msol and 0.99\msol, respectively. This stellar model was then relaxed to re-establish hydrostatic equilibrium \citep{Ohlmann2017}. For the relaxation process, we placed the star in a box with a size of $7\E{13}\,\textrm{cm}$ and a uniform background density of $\rho = 10^{-13}\,\mathrm{g\,cm}^{-3}$ and evolved it for ten dynamical timescales ($8.2\E{6}\, \textrm{s}$)\. . During the first half of the relaxation, we applied a damping force to remove spurious velocities; during the second half, the star was let free to evolve on its own, and we verified that the model remained hydrostatically stable.
We then simulated the common-envelope phase by placing the star and its companion in a box with a side length of $1.6\E{15}\,\mathrm{cm}$ ($2.35\E{4}$\rsol) at an initial separation of 49\rsol, which corresponds to the surface of the RG, with a corotation fraction of 95\%.

\subsubsection{CE merger}
As we cannot simulate the entire common-envelope phase with a fully resolved core and companion, we set up an entirely new simulation to study the GW signals released by a CE merger in which we considered only the part that generates the loudest highest-frequency GWs during the event: the merger of the red giant core and its companion. To ensure that the companion was not disrupted too far from the RG core, we chose a compact object; in particular, we took a 0.6\msol white dwarf (WD) as a companion for which 50\% of the mass was carbon and 50\% oxygen, and whose initial radius was 0.012 $R_\odot$. As first-order approximation, we modeled the RG core as a 0.4\msol WD made entirely of helium with a radius of 0.016\rsol.
The methods and setup of the white dwarfs in this simulation are similar to those of \cite{Pakmor2021}. The stellar models were constructed as simple hydrostatic equilibrium configurations assuming a uniform temperature of $2\E{7}\, \mathrm{K}$ for the RG core and $10^6 \, \mathrm{K}$ for the companion. We obtained their density profiles following the Helmholtz equation of state, and then we gave the \AREPO cells the appropriate masses. Both stars were given a seed magnetic field set up in a dipole configuration with a polar surface field strength of 10\,G.
The stars were relaxed separately to ensure hydrostatic equilibrium because mapping the initial 1D models to 3D can lead to discretization errors in the hydrostatic equilibrium. 
The relaxation process was the same for both stars: They were placed in a box of $10^{10}\,\mathrm{cm}$ with a static, uniform background grid density of $\rho = 10^{-5}\,\mathrm{g\,cm}^{-3}$. During the first half of the relaxation, we applied a damping force that reduced potential spurious velocities from discretization errors.
During the second half of the relaxation, the stars were left to evolve on their own (without the damping force), and we verified that the density profiles were stable. The duration of the relaxation was chosen to be ten times the dynamical timescale of each star,  that is,\  $20\, \mathrm{s}$ for the 0.6\msol WD and $63\, \mathrm{s}$ for the 0.4\msol RG core. 
After the relaxation, we placed the two relaxed stars into the same box with a box size of $7.3\E{11}\,\mathrm{cm}$ with a static, uniform background grid with $\rho = 10^{-5}\,\mathrm{g\,cm}^{-3}$. This background density was chosen to resemble the denser regions of a stellar envelope, but it has very little effect on the dynamics of the merger because at this stage, it will be driven by the mass transfer and tidal interaction between the core and the companion.
The mass resolution of the simulation was $m_\mathrm{cell} \approx 5\E{-7}$\msol. The initial distance between the WDs was chosen to be three times the separation at the point of Roche-lobe overflow, and then we artificially shrank the orbit until the onset of a steady mass transfer in a similar fashion as for the main-sequence star merger. This occurred at an orbital separation of $3.25\E{9}\,\textrm{cm}$.
At this point, we started the unmodified simulation of the merger.

\begin{table*}
\centering
\caption{\label{tab:setups}Summary of the simulations. a$_0$ is the initial orbital separation, a$_\mathrm{final}$  is defined as the separation between the stars right before disruption in the case of the MS star merger and the CE merger. In the case of the successful CE, as the orbit is eccentric, a$_\mathrm{final}$ is defined as the semi-major axis of the orbital separation between the core and the companion at the end of the simulation. $f_\mathrm{GW0}$ and $f_{\mathrm{GWfinal}}$  are computed as twice the orbital frequency assuming a Keplerian orbit and separation equal to a$_0$ and a$_\mathrm{final}$, respectively. In the case of the successful CE ejection, only the mass of the core and the companion are considered for the calculation of the Keplerian frequency.}
\begin{tabular}{c c c c c c c c} \toprule
 Model & $M_1$(\small{total}) & $M_1$(\small{core}) & $M_2$ & a$_0$ & $f_{\mathrm{GW,0}}$ & a$_{\mathrm{final}}$ & $f_{\mathrm{GW,final}}$\\
  & [$M_\odot$] & [$M_\odot$] & [$M_\odot$] & [$R_\odot$] & [Hz] & [$R_\odot$] & [Hz] \\
 [0.5ex]
 \midrule
 MS-merger & 9 & - & 8 & 6.2 &  $5.3\E{-5}$ & 5.0 & $7.3\E{-5}$ \\
 CE-merger & 0.63 & - & 0.42 & 0.047 &  $2.0\E{-2}$ & 0.039 & $2.5\E{-2}$ \\
 Successful CE & 2 & 0.37 & 0.99 & 49 & $6.8\E{-7}$ & 3.9 & $3.0\E{-5}$\\
 \bottomrule
\end{tabular}

\end{table*}

\subsection{Computing the gravitational waves}
To compute the gravitational wave signal, we calculated the approximate quadrupole radiation from Newtonian gravity. We followed the same approach as in  \cite{Seitenzahl2015} and used the method by \cite{Blanchet1990} and \cite{Nakamura1989} to compute the second derivative of the quadrupole moment. In this way, the gravitational quadrupole radiation field in the transverse-traceless gauge $\mathbf{h^\mathrm{TT}}$ takes the form
\begin{equation}
    h^\mathrm{TT}_{ij} (\mathbf{x},t) = \frac{2G}{c^4R}P_{ijkl}(\mathbf{n}) \int\dd^3x\,\rho\,(2v_kv_l-x_k\partial_l\Phi-x_l\partial_k\Phi).
\end{equation}
Here, $\mathbf{P}$ is the transverse-traceless projection operator,
\begin{equation}
\begin{split}
    P_{ijkl} (\mathbf{n}) = \,&(\delta_{ij} - n_in_k)(\delta_{jl} - n_jn_l) \\
                            &- \frac{1}{2}(\delta_{ij} - n_in_j)(\delta_{kl} - n_kn_l),
\end{split}
\end{equation}
with the normalized position vector $\mathbf{n} = \mathbf{x}/R$ and the distance to the source $R=|x|$. As usual, $\mathrm{G}$ is the gravitational constant, c is the speed of light, $\rho$ is the density, $\mathbf{v}$ is the velocity, $\partial_{i}$ represents a partial derivative with respect to the spatial coordinate $i,$ and $\Phi$ is the Newtonian gravitational potential. 

The amplitude of the GW radiation field can be written in terms of the two unit linear polarization tensors $\mathbf{e}_+$ and $\mathbf{e}_\times$, and when a line of sight is chosen, the amplitude of the GWs takes the form
\begin{equation}
    h^{\mathrm{TT}}_{ij}\,(\mathbf{x},t) = \frac{1}{R}(A_+ e^+_{ij} + A_\times e^\times_{ij}),
\end{equation}
where $A_+$ and $A_\times$ are the amplitudes of the two polarizations as a function of time. We worked in Cartesian coordinates and considered as line of sight the $z$-direction, corresponding to the initial axis of rotation of our binary systems. In this line of sight, the GW emission is strongest and therefore provides the most optimistic scenario in terms of detectability of the signal.
Along this axis, the polarization amplitudes take the form
\begin{subequations}
\begin{align}
    A^z_+ &= A_{xx} - A_{yy} \\
    A^z_\times &= 2A_{xy},
\end{align}
\end{subequations}
with $A_{ij}$ defined as
\begin{equation}
    A_{ij} = \frac{G}{c^4}\int\dd^3x\,\rho\,(2v_iv_j-x_i\partial_j\Phi-x_j\partial_i\Phi).
    \label{Aij}
\end{equation}
We implemented the computation of the terms $A_{xx}$, $A_{xy}$, $A_{xz}$, $A_{yy}$, $A_{yz}$ , and $A_{zz}$ in \AREPO following Eq.~(\ref{Aij}) in order to output the emitted GW signal on the fly during the simulations.

The characteristic strain of a GW signal is defined as $h_c(f) = 2f\Tilde{h}(f),$ where $\Tilde{h}$ is the Fourier transform of the time-domain amplitude of the GW,

\begin{equation}
    \Tilde{h}(f) = \frac{1}{R}\sqrt{|\Tilde{A}_+(f)|^2 + |\Tilde{A}_\times(f)|^2},
\end{equation}
with 
\begin{equation}
    \Tilde{A}(f) = \int^\infty_{-\infty}e^{-2\pi i f t}A(t)\dd t ,
\end{equation}
 where $A(t)$ is the amplitude of each polarization over time.
 We calculated these values using a discrete fast Fourier transform and a Tukey window \citep{Abbott2016a}. 
 To capture the main features of the signal at its start and end, we used an asymmetric Tukey window in the case of the successful CE ejection.
The signal-to-noise ratio (SNR) of a signal then takes the form \citep[e.g.,][]{Moore2015}
\begin{equation}\label{snr}
    (\mathrm{SNR})^2 = 4\int^\infty_0\frac{\abs{\Tilde{h}(f)}^2}{S_n(f)}\dd f ,
\end{equation}
where $S_\mathrm{n}$ is the noise power spectrum density of the detector. We took the approximate values for LISA from \cite{Robson2019}.

\section{Results\label{sec:results}}
\subsection{Main-sequence star merger\label{results:ms}}
\begin{figure}
\includegraphics[width=0.5\textwidth]{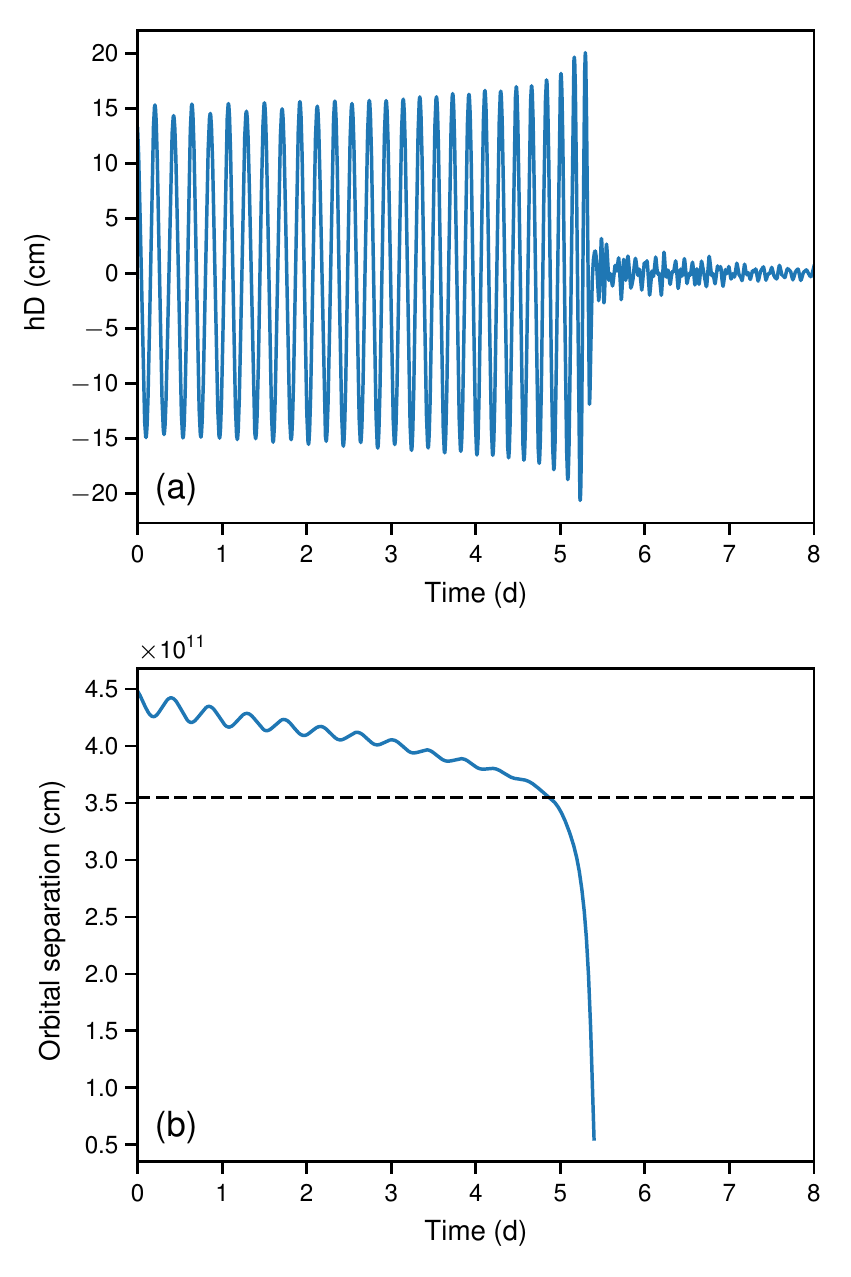}
\caption{\label{fig:ms-waveform}Evolution of the main-sequence star merger. Top (a): Strain of the GW signal $h$ over time (+ polarization in the $z$-direction, $h = A_+^z/D$). The amplitude is shown for an observer situated at a distance D.
Bottom (b): Orbital separation over time. The dashed black line indicates the last stable orbital separation before disruption at $3.5\E{11}\mathrm{cm}$.}
\end{figure}
\begin{figure}
\includegraphics[width=0.5\textwidth]{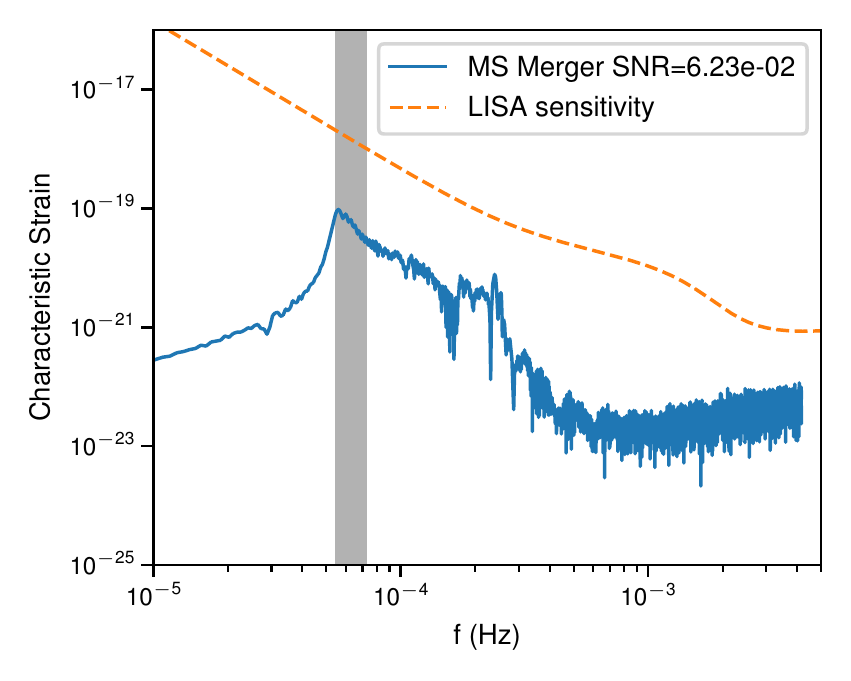}
\caption{\label{fig:ms-characteristic-strain}Characteristic strain of the GW signal released by the merger of two main-sequence stars (in blue) assuming a distance of $1\,$kpc to the source along the polar axis. The sensitivity of LISA from \cite{Robson2019} is shown by the dashed orange line. The region shaded in gray represents the frequency range spanning between twice the orbital frequency at the beginning of the simulation and twice the orbital frequency at the moment the star is disrupted (assuming the orbits to be perfectly circular and Keplerian).}
\end{figure}

A full description of the hydrodynamic outcome of the merger between the main-sequence stars can be found in \cite{Schneider2019}.
The initially more massive (primary) star transfers mass to the initially less massive (secondary) star for a few days, draining angular momentum and shrinking the orbit. By the time the orbital separation reaches the tidal disruption radius of the primary star, commonly approximated as
\begin{equation}
    r_t = \left(\frac{M_2}{M_1}\right)^{1/3}R_1,
    \label{eq:tidal_approx}
\end{equation}
where $M_1$ and $M_2$ are the masses of the primary and the secondary stars and $R_1$ is the radius of the disrupted star (primary), it is disrupted, and its central material merges with that of the secondary star.
The resulting remnant consists of a $\sim 14$\msol central core with a $3$\msol torus surrounding it.
The gravitational waves determined from the 3D MHD simulation of the merger event are shown in Figure \ref{fig:ms-waveform}.

The amplitude of the strain increases as the orbital separation shrinks, and it peaks at the moment of the disruption of the primary star. Immediately after the merger takes place, the emission ceases. This decline in signal strength is expected as the remnant rapidly evolves to a mostly axisymmetric configuration.
The characteristic strain of the GWs is shown in Figure \ref{fig:ms-characteristic-strain} assuming a distance of $1\,$kpc between source and detector. For this distance, we obtain an SNR of $0.062$ with LISA. The peak of the characteristic strain is found at a frequency of about $5.4\E{-5}\,\mathrm{Hz,}$
which corresponds to an orbital frequency of $2.7\E{-5}\,\mathrm{Hz}$. 

The Keplerian orbital separation for a system orbiting at this frequency is $4.3\E{11}\,\mathrm{cm}$.
This value is very similar to the separation between the stars at the beginning of our computations. The characteristic strain integrates the signal over the observation time, so that the longer the system spends at a certain orbital frequency, the higher the associated characteristic strain for that frequency. In our simulation, the binary system only evolves through a small range of orbital separations before the disruption between $4.3\E{11}\,\mathrm{cm}$ and $3.5\E{11}\,\mathrm{cm}$, which  corresponds to a narrow GW frequency range from $5.4\E{-5}\,\mathrm{Hz}$ to $7.3\E{-5}\,\mathrm{Hz}$ (shaded in gray in Figure~\ref{fig:ms-characteristic-strain}). Theoretically, we expect most of the GWs produced in our simulation to be within this frequency range.
We lack simulation data at larger separations, which explains the cutoff at frequencies below this range. We could also have lower frequencies from ejected material after the disruption, so the characteristic strain is not zero.
The signal at higher frequencies (smaller separations) originates from the phase after disruption of the primary star, and hence, we see a drop-off in the characteristic strain at frequencies higher than the frequenciy corresponding to the separation equal to the tidal disruption radius of the star. 
The characteristic strain at these higher frequencies ($f_{\mathrm{GW}}>10^{-4}\,\mathrm{Hz}$) comes from the weak post-merger signal.

It is difficult to compute the tidal disruption radii in noncompact interacting systems. The usual approximation for the tidal disruption radius, shown in Eq.~(\ref{eq:tidal_approx}), is obtained by equating the self-gravity force of the disrupted star with the gravitational pull that it experiences from the companion object and assuming that the radius of the star is much smaller than the separation between the stars. 
Therefore, this approximation does not hold true when the tidal disruption radius is on the order of magnitude of the radius of the star.

To obtain a better approximation, we calculated the tidal force across the disrupted star. First, we computed the gravitational acceleration at the points of the star that are closest and farthest from the companion:
\begin{equation}
    a_\mathrm{grav} = \frac{GM_1}{(d\pm R_2)^2}
,\end{equation}
where $d$ is the distance to the companion. To obtain the tidal acceleration, we subtracted the contribution from the acceleration on the near point from the contribution from the point far out,
\begin{equation}\label{tidal_acc}
\begin{aligned}
    a_\mathrm{tidal} & = GM_1\left[\frac{1}{(d-R_2)^2}-\frac{1}{(d+R_2)^2}\right]  \\ 
    & = GM_1\frac{4dR_2}{(d^2-R_2^2)^2}.
\end{aligned}
\end{equation}
The acceleration due to self-gravity is 
\begin{equation}
    a_{\mathrm{self}} = \frac{GM_2}{R_2^2}.
    \label{self-grav}
\end{equation}
At the point of tidal disruption, $d = r_t$ and $a_\mathrm{tidal} = a_\mathrm{self}$. Therefore, equating  Eq.~(\ref{tidal_acc}) and Eq.~(\ref{self-grav}), we obtain
\begin{equation}
\label{tidal_radius}
    \frac{M_2}{M_1} = \frac{4R_2^3 r_t}{(r_t^2 - R_2^2)^2}.
\end{equation}
Solving Eq.~(\ref{tidal_radius}) numerically for the tidal disruption radius, we obtain a better approximation of its value. In the following, we define the tidal disruption radius as the solution of Eq.~(\ref{tidal_radius}) unless stated otherwise.

Another source of uncertainty is the fact that the stellar radius is difficult to define during ongoing mass transfer.
For example, using the original radius of the star from the MESA models, we obtain a tidal disruption radius of $5.5\E{11}\,\mathrm{cm}$, which is larger than the initial separation of our system. When we consider as the stellar radius the distance from the center in which 99\% of the stellar mass is contained, we obtain a more sensible value for the tidal disruption radius of $3.5\E{11}\,\mathrm{cm}$. Figure \ref{fig:ms-waveform} shows that the latter value matches the actual point of disruption well.
With this, we aim to show that the tidal disruption radius of a star in a binary system can give us a rough estimate of the highest frequency of the GWs that the merger would release during the inspiral phase.

\begin{figure}
\includegraphics[width=0.5\textwidth]{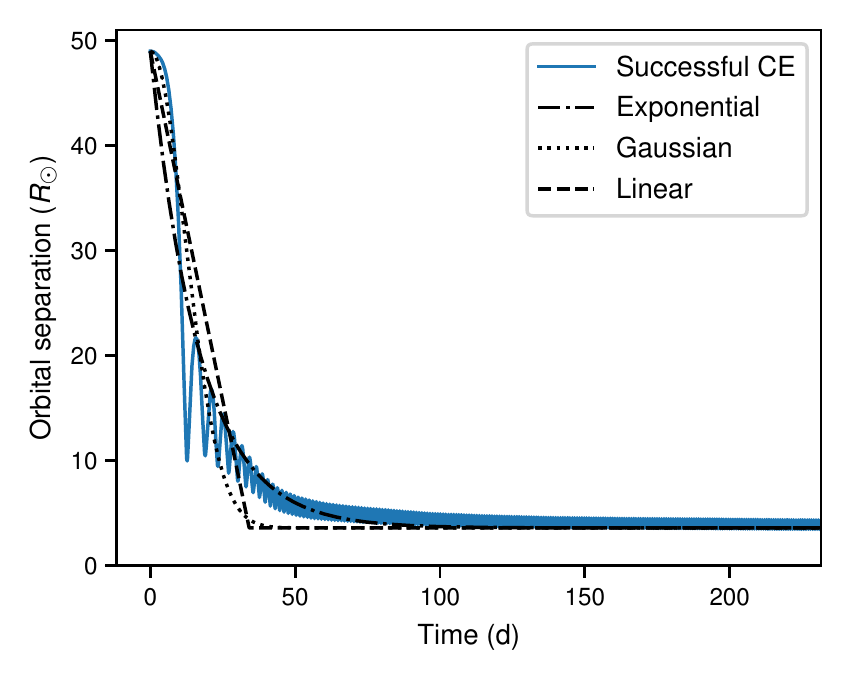}
\caption{\label{fig:point-mass} Time evolution of the orbital separation during the successful CE ejection. In blue we show the data from the \AREPO simulation. The black lines show the fit to these data with an exponential, a Gaussian, and a linear orbital decrease.}
\end{figure}

\subsection{\label{results:succesful}Successful common-envelope ejection}
\begin{figure}
\includegraphics[width=0.5\textwidth]{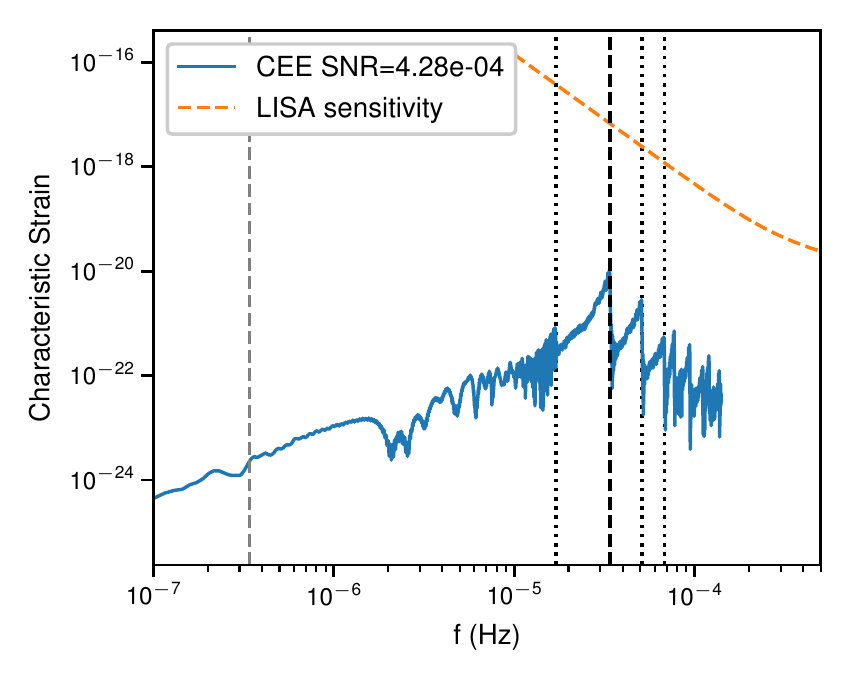}
\caption{\label{fig:ce-characteristic-strain} Characteristic strain of the GW signal released by the successful common-envelope ejection (in blue) assuming a distance of 1\,kpc to the source. The sensitivity of LISA from \cite{Robson2019} is shown by the dashed orange line. The vertical dashed gray line represents twice the orbital frequency at the beginning of the CE (assuming a perfectly circular and Keplerian motion). The dashed black line represents twice the orbital frequency after the ejection of the envelope. The dotted black lines represent the first four harmonics of the final orbital frequency ($f_n=n\cdot f_\mathrm{orb,final}$).}
\end{figure}
The initial separation between the core of a giant star and its companion is generally too large for the detection of the system with LISA. The frequencies of the associated GWs are too low because the radii of giant stars are on the order of 10 to 100 \rsol. In order for these events to be detectable, the orbit must shrink to a separation at which the orbital frequency is in the appropriate frequency band for LISA.
The evolution of our simulation of a successful CE ejection is very similar to that of \cite{Sand2020}, but with an RG instead of an asymptotic giant branch (AGB) star, and the whole interaction occurs on a shorter timescale.

As soon as the simulation starts, the companion plunges into the envelope and its orbit shrinks for a few hundred days, until it ejects most of the envelope mass. From this point on, there is little orbital evolution on the simulated timescale, and the binary remains on an eccentric orbit with a semi-major axis of 3.6\rsol after 200 days. The evolution of the orbital separation is shown in Fig.~\ref{fig:point-mass}. At this final separation, the companion orbits the core of the RG with a frequency of 0.017\,mHz. In contrast to the other two simulations, the GW signal is not shut down. Instead, the main component of the signal evolves from low to higher frequencies.
The characteristic strain of this simulation is shown in Fig.~\ref{fig:ce-characteristic-strain}. It peaks at a frequency of $3.4\E{-5}\,\mathrm{Hz}$, which corresponds to exactly twice the orbital frequency after the envelope has been ejected. The final orbit is not perfectly circular; it has an eccentricity of 0.1. Keplerian elliptic orbits emit GWs at multiple harmonics (frequencies that are integer multiples) of the orbital frequency \citep{Maggiore2008}. We observe this in the characteristic strain of the signal as further peaks located at frequencies equal to the harmonics of the final orbital frequency. The frequency of maximum strain is also still too far from the peak sensitivity of LISA to be detectable; we obtain an SNR of $4.28\E{-4}$.

\begin{figure}
\includegraphics[width=0.5\textwidth]{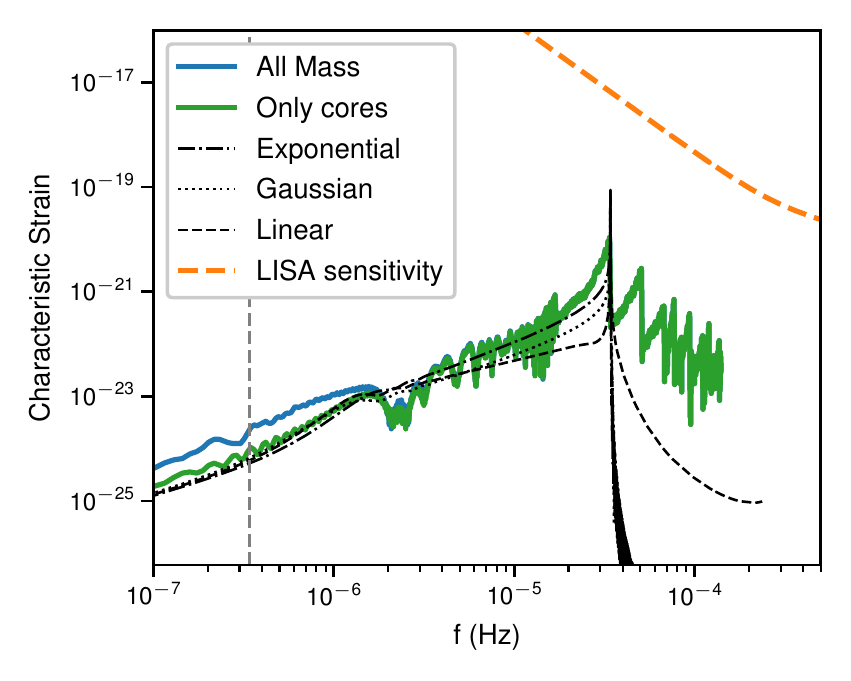}
\caption{\label{fig:envelope} Characteristic strain of the successful CE ejection. The characteristic strain of the GWs released by the entire simulation (the core of the RG, its envelope, and the companion) is shown in blue. In green we show the characteristic strain of the GW signal produced by only the core of the RG and the companion. The dashed gray line represents twice the orbital frequency at the beginning of the simulation. The black lines show the characteristic strain produced by toy models (see Sect.~\ref{subsec:toymodel}). The  sensitivity of LISA from \cite{Robson2019} is shown by the dashed orange line. }
\end{figure}

A common approximation used for calculating the GWs generated by CEs is to only consider the core of the RG and the companion as point-mass particles and neglect the effect of the envelope on the GW signal \citep{Ginat2020, Renzo2021}. With our simulation, we can test how accurate this treatment is. Figure \ref{fig:envelope} shows the difference in the characteristic strain of the GWs released by the entire system and the strain resulting from the cores alone. The characteristic strain differs slightly at the lowest frequencies, but is essentially the same at frequencies higher than $3\E{-6}\,\mathrm{Hz}$. As most of the envelope mass is ejected early on in the simulation when the orbital frequency of the system increases rapidly, the envelope is not expected to contribute  significantly at GW frequencies much higher than twice the initial orbital frequency. 
As the system moves to higher orbital frequencies by ejecting the envelope, the contribution to the GW signal from the envelope necessarily decreases at higher frequencies. The difference between the two characteristic strains should therefore mostly be present at GW frequencies close to the one corresponding to the initial orbital frequency. In any case, the magnitude of this discrepancy is very small, therefore we can safely state that the approximation holds true for these systems in regard to their detectability with LISA.
It is not clear whether this also applies to a CE evolution that leads to a merger. If the envelope is not completely ejected, it may contribute to the GW signal at higher frequencies.
Furthermore, while there is no significant direct contribution of the envelope to the formation of the GW signal, an accurate representation of the core-envelope interaction in the simulations is still necessary to reliably determine the rate of inspiral and the final orbital separation between the cores that ultimately shape the GW signal.

\subsection{\label{results:failed-ce}Common-envelope merger}
\begin{figure*}
    \centering
    \includegraphics[width=\textwidth]{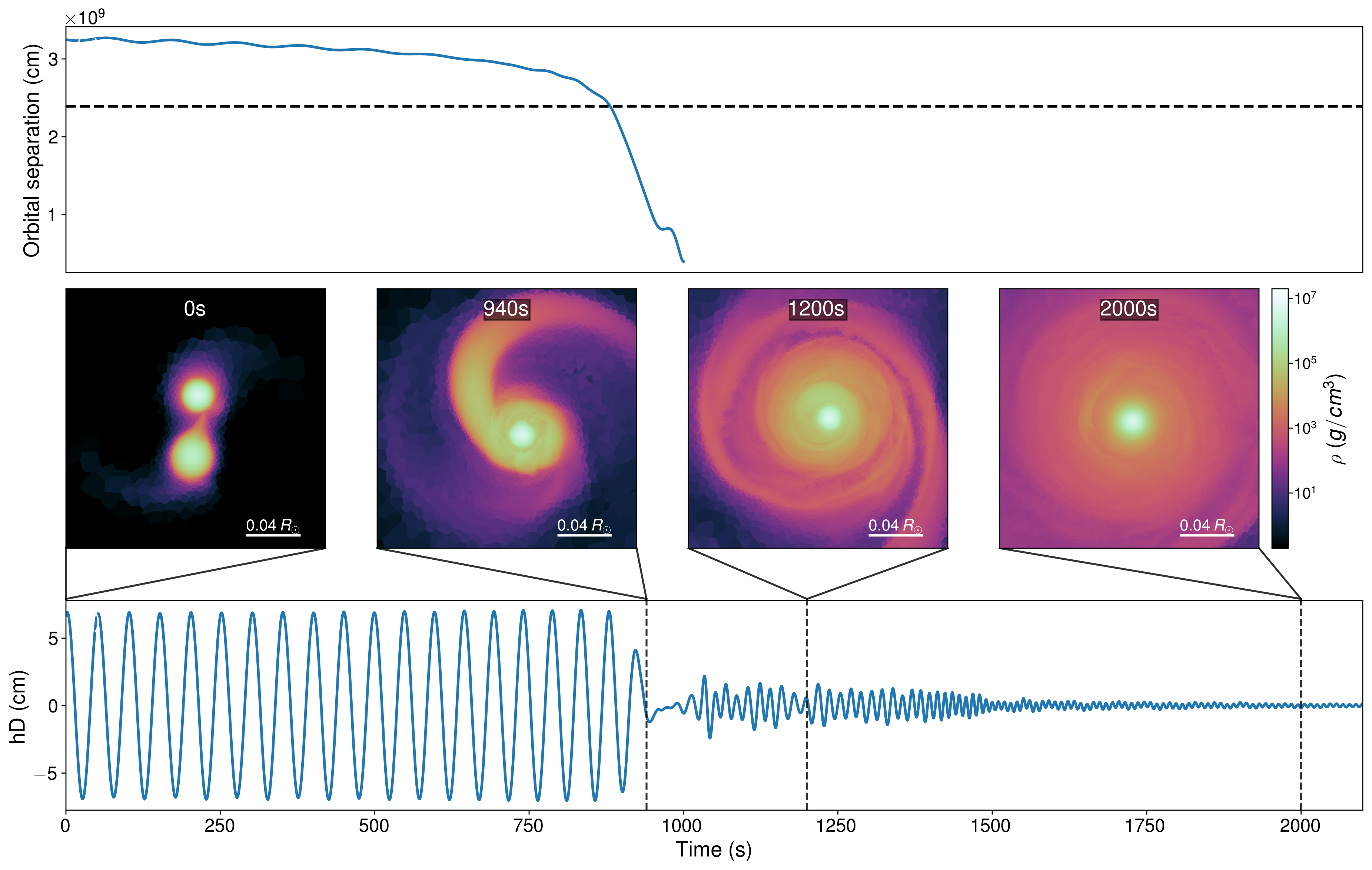}
    \caption{
    Evolution of the CE merger.
    \emph{Top}: Orbital separation over time. The dashed black line indicates the computed tidal disruption radius at $2.4\E{9}\mathrm{cm}$.
    \emph{Middle}: Hydrodynamical evolution of the core merger resulting from a CE merger. The density is color-coded. The computation of the gravitational waves starts at the onset of steady mass transfer. The more massive companion shreds mass from the less massive, less dense core of the RG. The separation between the two objects slowly decreases until it reaches the point of tidal disruption at $\sim$ $840\,$s. The companion accretes a fraction of the mass from the former RG core ($\sim$ 0.11\msol), while the remaining mass forms a disk-like structure around it.
    \emph{Bottom}: Strain of the GW signal over time released by the CE merger (cross-polarization in the $z$-direction, $h = A_+^z/D$).
}
    \label{fig:hydro_failedCE}
\end{figure*}
\begin{figure}
\includegraphics[width=0.5\textwidth]{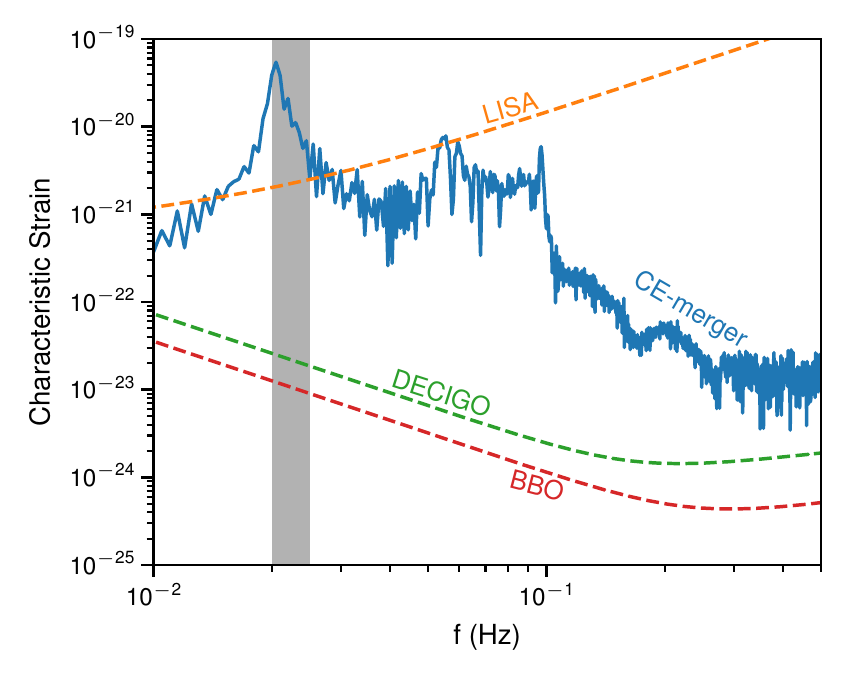}
\caption{\label{fig:failedcee-characteristic-strain}Characteristic strain of the GW signal of the CE merger (in blue) assuming a polar distance of 1\,kpc to the source. The sensitivities of LISA \citep{Robson2019}, DECIGO, and BBO \citep{Yagi2011} are shown as  dashed orange, green, and red lines, respectively. The region shaded in gray represents the frequency range spanning between twice the orbital frequency at the beginning of the simulation and twice the orbital frequency at the moment when the star is disrupted.}
\end{figure}
\begin{figure}
\includegraphics[width=0.5\textwidth]{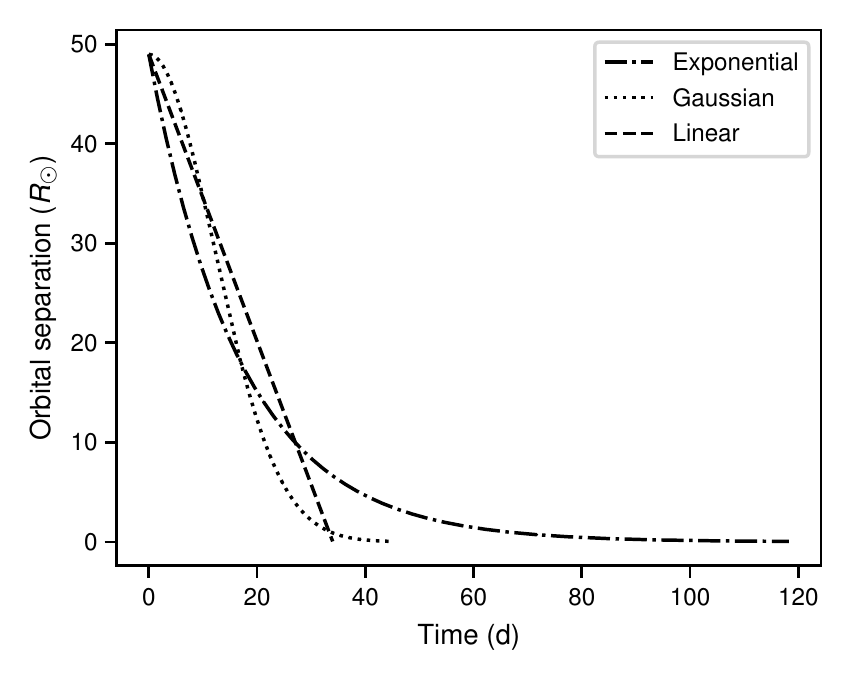}
\caption{\label{fig:failedcee-separation-inspiral}Time evolution of the orbital separation of our toy models. The orbit shrinks from the initial orbital separation of the successful CE simulation to the initial orbital separation of the CE merger.}
\end{figure}
\begin{figure}
\includegraphics[width=0.5\textwidth]{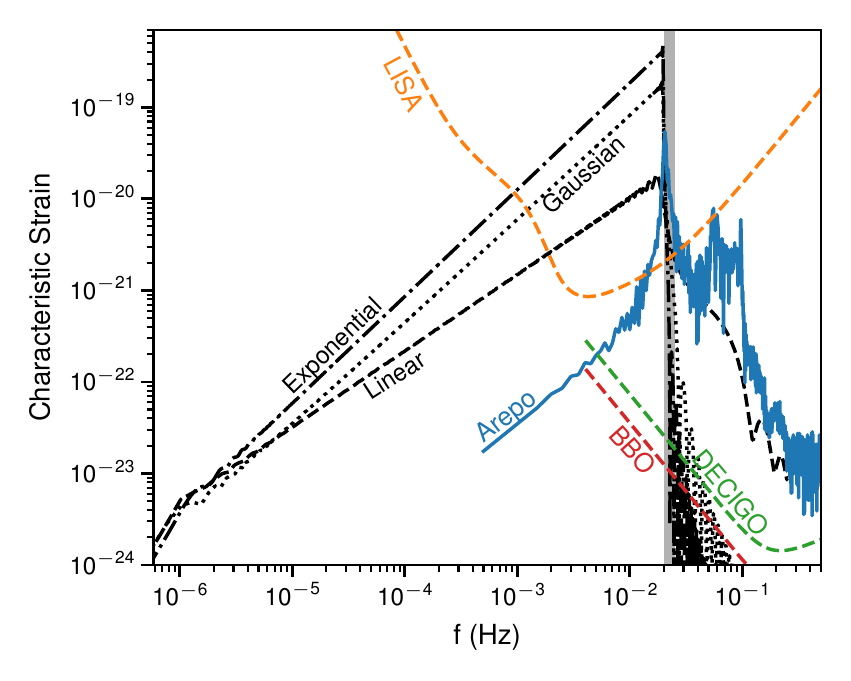}
\caption{\label{fig:failedcee-inspiral}Characteristic strain of the GW signal released by the CE merger (in blue) assuming a distance of 1\,kpc to the source along the polar axis. The sensitivities of LISA \citep{Robson2019}, DECIGO, and BBO \citep{Yagi2011} are shown as dashed orange, green and red lines, respectively. 
The region shaded in gray represents the frequency range spanning between twice the orbital frequency at the beginning of the simulation and twice the orbital frequency at the moment when the star is disrupted (assuming the orbits to be perfectly circular and Keplerian).
The black lines show the characteristic strain produced by our toy models.}
\end{figure}
The hydrodynamical evolution of the CE merger is summarized in Figure \ref{fig:hydro_failedCE}. 
From the beginning of the simulation, the less massive, less dense core of the RG primary star transfers mass onto the more massive and more compact companion. The orbit shrinks for a few hundred seconds due to this mass transfer until it reaches the point of tidal disruption and the RG core is disrupted around the inspiraling WD companion.
The companion accretes a fraction of the mass of the former RG core, approximately $0.11$\msol.
The remaining mass forms a disk-like structure around the remnant. This structure is not axisymmetric immediately after disruption, but it becomes so after a few hundred seconds.

The strain of the GWs (Fig. \ref{fig:hydro_failedCE}, bottom) is quite similar to the emission of the merger between MS stars, with a few differences. 
For the WDs, the merger occurs on a shorter timescale (hundreds of seconds instead of days). 
The amplitude of the GWs does not noticeably increase immediately before the merger, but its maximum value rather remains mostly constant until the disruption.
Moreover, the post-merger signal is initially more pronounced than in the case of the MS star merger. The explanation for this stronger post-merger signal is the initial lack of axial symmetry in the remnant. As the disk homogenizes, the signal declines.

There is again a narrow range of orbital separations before the binary is disrupted. Our initial separation is $3.25\E{9}\,\textrm{cm}$, which corresponds to a GW frequency of $2\E{-2}\,\textrm{Hz}$. No significant contribution to the signal is expected at frequencies lower than this value in our simulation.
The computation of the tidal disruption radius for this simulation is even more challenging than in the case of the MS star merger because the disrupted star is a WD. This means that its radius will increase as it loses mass. Eq.~(\ref{tidal_radius}) shows that the tidal disruption radius increases with the stellar radius. Therefore, the tidal disruption radius is underestimated when  the initial value of the radius of the RG core is used. When the distance from the center in which 99\% of the stellar mass is contained is used as the radius, we find $r_t = 2.4\E{9}\,\textrm{cm }$following Eq.~(\ref{tidal_radius}), which corresponds to a GW frequency of $3.17\E{-2}\,\textrm{Hz}$, assuming the orbit to be perfectly circular and Keplerian. The top panel of Figure~\ref{fig:hydro_failedCE} shows that this value matches the point of disruption well.

Computing the characteristic strain (Figure \ref{fig:failedcee-characteristic-strain}), we find that it peaks at a frequency of $2\E{-2}\,\mathrm{Hz}$, which matches twice the orbital frequency at the beginning of the simulation. This is expected because this is the largest orbital separation that we simulated, therefore we do not predict any significant GW signal at lower frequency. In this frequency range, LISA is far more sensitive than in the case of the MS star merger.
For this reason, the characteristic strain of the CE merger ejection peaks above the sensitivity curve of LISA for a signal that lasts only a few minutes, generating an SNR of $6.6$ at a distance of 1\,kpc.
The post-merger signal also contributes some high-frequency components to the characteristic strain. There is some relatively high strain up to frequencies close to $0.1 \,\textrm{Hz}$.
These higher frequencies are in the correct range for future decihertz GW observatories such as DECIGO or BBO. We computed the SNR for the CE merger simulation with DECIGO and BBO using the sensitivity curves described in \cite{Yagi2011}, and we obtained values of 896 and 1865, respectively, at a distance of 1\,kpc. If designed according to current plans, these instruments would be able to detect the final disruption of CE-mergers with a much higher SNR due to the increase in sensitivity and the higher frequency band.

The computed value of the tidal disruption radius has been underestimated and thus overestimates the highest frequencies released by the binary before the merger (end of the region shaded in gray in Figure~\ref{fig:failedcee-characteristic-strain}). It is still within a factor of two from the peak frequency, however, and can be used as an upper bound.

\subsection{Toy model for CE inspiral\label{subsec:toymodel}}
Based on our successful CE ejection simulation, we modeled the plunge-in of the companion inside the red giant envelope with simple analytical formulae in order to determine how the characteristic strain obtained from them compares with the strain of our simulation.
We used three functional forms to model the orbital decay: an exponential, a Gaussian, and a linear function, defined as
\begin{subequations}\label{eq:toy_models}
\begin{align}
   &a_\mathrm{exp}(t) = (a_0-a_{\mathrm{final}})\,e^{-\frac{t}{\tau}} + a_\mathrm{final}, \\
   &a_\mathrm{gauss}(t) = (a_0-a_{\mathrm{final}})\,e^{-\left(\frac{t}{\tau}\right)^2} + a_\mathrm{final}, \\
   &a_\mathrm{lin}(t) = \begin{cases}
       (a_0-a_{\mathrm{final}})\left(1-\frac{t}{\tau}\right)+a_\mathrm{final}&  t \leq \tau,\\
       a_\mathrm{final}& t > \tau.
    \end{cases}
\end{align}
\end{subequations}
Here $a_0$ and $a_\mathrm{final}$ have the values specified in Table~\ref{tab:setups}, and $\tau$ in each formula was chosen to be the timescale of the orbital decay such that it fits the rate observed in our successful CE simulation. We find $\tau = 1.47\times 10^6$s for the exponential and Gaussian models, and $\tau = 2.94\times 10^6$s for the linear model.
We followed these models for a time equal to the longest time in our simulation.
Figure \ref{fig:point-mass} shows that the Gaussian model captures the early part of the plunge-in best, whereas the exponential model matches it better at a later stage. In these simple models, we assumed no eccentricity, such that there is a smooth decrease of the orbital separation.
We can then compute the GWs that would be released by two point-masses following the orbital evolution as described in \cite{Creighton},
\begin{subequations}
\begin{align}
   &A_+(t) =  \frac{4G\mu}{c^2}\left(\frac{\nu(t)}{c}\right)^2 \mathrm{cos}\,\phi(t) \label{eq:toy_gw_1}\\
   &A_\times(t)  =\frac{4G\mu}{c^2}\left(\frac{\nu(t)}{c}\right)^2 \mathrm{sin}\,\phi(t), \label{eq:toy_gw_2}
\end{align}
\end{subequations}
where $\mu$ is the reduced mass, $\nu(t)$ is defined as $\sqrt{\frac{GM_\mathrm{total}}{a(t)}}$ and $\phi(t)$ is the phase of the GWs, which we obtained by numerically solving $\Dot{\phi}=2\pi f_\textrm{GW}(t) = 4\pi f_\mathrm{orb}(t)$. For these toy models, we assumed that the point-masses are always on a Keplerian circular orbit in order to obtain the instantaneous value of the orbital frequency. 
We calculated the characteristic strain of these toy models and compare it with the data from our simulations in Figure \ref{fig:envelope}. As we explicitly specified a final orbit, we have a sharp peak at twice the final orbital frequency of the system, and no features at its harmonics as there is no eccentricity. In general, the shape of the strain of the toy model GWs reasonably matches the strain of the \AREPO simulation. For the lowest frequencies produced by our simulation, the three toy models give the same value for the GW strain as in the hydro simulation when only the core masses were taken into account. For frequencies greater than $3\E{-6}\,\mathrm{Hz}$, the exponential model reproduces the shape of the strain in the simulation best. The three toy models produce an SNR similar to the SNR in the simulation ($4.00\E{-4}$ for the exponential, $4.16\E{-4}$ for the Gaussian model, and $4.17\E{-4}$ for the linear model). We conclude from this that the models capture the strain of the GWs in the specified frequency range moderately well and can be used as simple toy models for the inspiral of our CE simulation.

Assuming that this approach also holds valid in a CE merger scenario, we applied it to the phase leading up to our CE merger simulation in order to estimate the GW signal produced during its inspiral phase.
Disregarding any potential signal from the envelope, we took two point mass particles with the masses of our WD and the core of the RG to compute the reduced mass in Eqs.~(\ref{eq:toy_gw_1}) and (\ref{eq:toy_gw_2}) and shrank the orbit according to Eq.~(\ref{eq:toy_models}) at a rate that resembles what we observe in our accurate simulation of the successful CE ejection ($\tau = 1.47\times 10^6$s in the exponential and Gaussian models, and $\tau = 2.94\times 10^6$s for the linear model). However, instead of stopping the inspiral at a distance of $3.6$\rsol, we let the orbit shrink to a separation of $a_\mathrm{final} = 3.25\E{9}\,\textrm{cm}$, equal to the initial orbit used in our simulation of the CE merger. With these toy models, we obtained an estimate of the characteristic strain of the GWs released during the earlier stages of a CE merger ejection. 
The orbital evolution of our toy models is shown in Fig.~\ref{fig:failedcee-separation-inspiral}. 

As observed in our previous simulations, the orbital frequency shifts from lower values up to the initial values of the CE merger simulation, while the strength of the GWs increases. The duration of this inspiral phase differs between our toy models, as we stopped them the moment they reached a separation equal to the initial orbital separation in our core merger simulation. The more time the system spends orbiting at a certain orbital frequency, the greater the strain of the GWs associated with that frequency.
The exponential model has the highest value for the characteristic strain at high frequencies because it spends more time at orbital separations close to the  desired final orbit, and the linear model has the lowest SNR because it reaches its final orbit faster.
The signal during this stage crosses the peak sensitivity of LISA, and therefore, most of the SNR for this observatory could actually come from this phase depending on the exact rate of the orbital decay. With the chosen toy models, we indeed obtain SNR values for the inspiral phase of 191.1 for the exponential model, 84.5 for the Gaussian, and 10.5 for the linear model (to compute the SNR in these toy models, we ignored the characteristic strain at frequencies higher than twice the orbital frequency at the final orbital separation because this part of the characteristic strain is just numerical noise introduced by the fast Fourier transform on nonperiodic data). The final part of the inspiral phase begins to enter the frequency range of  DECIGO and BBO, and because of the planned increase in sensitivity with respect to LISA, they would be even better able to measure this final stage of the CE merger, obtaining DECIGO SNRs of 6784, 2961, and 299 and BBO obtaining SNRs of 13957, 6092, and 616 for the exponential, Gaussian, and linear models, respectively.
\section{Discussion\label{sec:discussion}}
We computed the SNRs for all our simulations following Eq.\,(\ref{snr}), and assuming a polar distance to the source of 1\,kpc. This distance is small and provides a best-case scenario for the detection of these events. The SNR is inversely proportional to the distance between detector and source, that is, if the event were to take place at 10\,kpc rather than 1\,kpc, the SNR would be ten times lower.

In the case of the merger between two MS stars, we obtain an SNR of 0.062, which suggests that these events are not suitable candidates for LISA sources. In Figure ~\ref{fig:ms-characteristic-strain} we show that the characteristic frequency of the GWs of this type of mergers is too far from the peak sensitivity of LISA. We started the simulation only a few days prior to the merger, whereas LISA could theoretically observe such a system for a few years before the disruption takes place, building up more signal and producing a higher SNR, but the frequency of these GWs will still be in the low-sensitivity band. The strength of the produced signal would increase in the case of a more massive merger \citep[GW amplitude $\propto M_{\textrm{chirp}}^{5/4}$;][]{Creighton}, where $M_\textrm{chirp}$ is the chirp mass of the binary,
\begin{align}
    M_{\textrm{chirp}} = \frac{(m_1m_2)^{3/5}}{(m_1+m_2)^{1/5}}.
\end{align}
However, it would also likely mean a larger tidal disruption radius, leading to lower GW frequencies and a decreased LISA sensitivity. The maximum frequency of GWs released by these systems before disruption can be estimated as twice the orbital frequency at the point where the separation between the stars equals the tidal disruption radius. We conclude that even at higher masses or longer observing times, the mismatch between the frequency of the GWs and the peak sensitivity of LISA makes main-sequence star mergers an unlikely GW source for this observatory.

Figure \ref{fig:ce-characteristic-strain} shows that in our simulation of a successful CE ejection, the final orbit reached is not tight enough to produce GWs in a frequency that LISA can detect even at an optimistic distance of $1\,$kpc. As indicated by \citealt{Renzo2021}, the final orbit resulting from a CE ejection has to shrink below one solar radius to be in a frequency range where LISA has a chance to detect them. Unfortunately, we do not observe a sufficient orbital shrinkage in this simulation or in any of the other simulations resulting in successful CE ejection that we analyzed \citep{Sand2020,Ondratschek2022,Moreno2021}. Therefore, we do not obtain any detectable GW signal from the dynamical phase of successful CE ejections for the LISA mission. Our final orbital separations are in line with those from \cite{Iaconi2019}, showing that most simulations of successful CE ejections produce final orbits larger than 1\rsol.

Different setups might shrink the orbit further.
All our setups involved giant stars with loosely bound envelopes that require less energy to be transferred from the binary orbit to eject them. It is possible that CE events in which the giant star has a more strongly bound envelope might result in tighter final orbits. There are some indications that the final orbits reached in our simulations are too large compared to observations \citep{Iaconi2019}. Many of the close binaries used as LISA verification targets \citep{Stroeer2006,Kupfer2018} will have evolved through a successful CE phase. It is not clear whether their orbits shrink to the current tight orbits after the CE or in the dynamical phase of the CE.
This further shrinking stage after the dynamical plunge-in could be due to a self-regulated inspiral, as suggested by \cite{Ivanova2013} or other mechanism draining angular momentum from the binary orbit (e.g., a circumbinary disk) such
that the GWs are in the frequency band of LISA.

Figure \ref{fig:failedcee-characteristic-strain} showed that the frequencies of the GWs in our CE-merger simulation are in a band in which LISA is more sensitive than in the case of the merger between two main-sequence stars. Computing the SNR, we obtain a value of 6.6, even though the total simulated time in this case is only a few thousand seconds. The frequency of the post-merger signal of this simulation is close to the optimal frequency range of decihertz GW observatories such as DECIGO and BBO, and the signal from our merger would produce SNRs of 896 and 1865, much higher than LISA due to the much higher planned sensitivity of the detectors, and because their frequency bands are able to best detect the post-merger signal.
In Fig.~\ref{fig:failedcee-inspiral} we showed that a large fraction of the GW signal is likely to be produced during the inspiral phase: As the orbital separation decreases, the frequency of the GWs will move from low frequencies (where LISA is not very sensitive) up to higher frequencies, crossing the peak sensitivity of LISA. 
The exact rate of the orbital decay strongly impacts the SNR, and at the same time, the dynamical plunge-in is one of the lesser known stages of the CE phase. A detection of such an event could provide information about stellar interiors in RG stars and the drag forces that the companion experiences while spiraling into the envelope. For example, different density profiles of the star lead to differences in the temporal evolution of the drag forces, causing very different waveforms and SNRs. Even if we were to observe the electromagnetic counterpart but do not detect the GW signal, a slow inspiral might be ruled out and a rapid plunge-in through the envelope might be indicated.
Therefore, accurate models for this inspiral stage are required for generating waveform templates and estimating the detection rates. In turn, future observations may reveal binary dynamics.

Using semi-analytical models and smoothed particle hydrodynamics simulations, \cite{Ginat2020} computed the waveform and characteristic strain of the inspiral phase of CE mergers. Their simulations covered the initial phase of the CE stage, but their models were stopped at larger orbital separations than the initial separation in our simulation, and a core merger was assumed to take place. The authors showed that the GWs released until that point might be observable by LISA.
In our 3D models, we followed the final disruption in detail, probing down to shorter orbital separation (higher frequencies), and our findings confirm that statement. The disruption on its own already produces a strong signal that could be detected if the source were close enough to LISA. Although we did not compute a detailed model of the late inspiral phase, our estimates using toy models of the inspiral point in the same direction as those of \cite{Ginat2020}.  With SNRs ranging between approximately 10 and 191 (depending on the model), the signal from the entire CE event could be strong enough for a detection with LISA. 
We also computed the SNR for the signals of these toy models with DECIGO and BBO. If their final design achieves the planned sensitivity, they might be able to detect the final phase of the inspiral in CE mergers with a much higher SNR than LISA, ranging between approximately 300 and 7000 for DECIGO and between 600 and 14000 for BBO, allowing them to be found at much larger distances.

Our findings about the detectability of these events with both LISA and the decihertz observatories are expected to hold for different masses  for the core of the giant and also the inspiraling companion.
Depending on the exact masses, we might enter a regime in which the CE merger of two white dwarfs leads to runaway thermonuclear burning and to an explosive event \citep{Pakmor2021a}, which, in contrast to type Ia supernovae, is expected to contain substantial amounts of H and He in its ejecta \citep[this corresponds to the so-called core degenerate supernova scenario of][]{Soker2011}. However, we did not include nuclear burning here.
In the event that one of the stars explodes, the GW signal will diverge from our results from that point on. The explosion could release a strong GW signal \citep{Seitenzahl2015} and the emission would cease immediately thereafter. If the explosion occurred during or after the disruption of the core, mainly the high-frequency components of the GWs will be affected, which does not alter the SNR significantly, and makes the system an interesting candidate for a multimessenger detection.

Changing the mass of the companion changes the tidal disruption radius of the core, and thus the highest GW frequency reached by the system before the disruption. A higher-mass companion such as a neutron star will disrupt the core at a larger distance and will also produce a stronger signal because the GW amplitude is proportional to $M_{\textrm{chirp}}^{5/4}$, but the maximum frequency is more sensitive to the core radius and mass than to the companion mass. 

If the companion were to be a neutron star, the formation of a Thorne-\.Zytkow object might be observed \citep[][]{Thorne1977,Hirai2022} because the strength and frequency of the GW signal during the inspiral and merger would not sensibly differ from the one we computed here. In particular, these objects are predicted to show high-frequency features in the post-merger signal, for whose detection decihertz observatories would be very well suited.

\section{Conclusions\label{sec:conclusion}}
We have computed the GW signal released during the merger between two MS stars, a successful CE ejection, and a CE merger in order to study their detectability with LISA. The simulations were carried out with the 3D MHD code \AREPO.

The GWs from mergers between main-sequence stars are found to be outside the sensitivity range of LISA at distances equal to or larger than $1\,$kpc. This result holds true for every merger between main-sequence stars, regardless of their masses.
The tidal disruption radii of stars in binary systems can be used to approximate the cutoff frequency of the GWs released by these systems during their mergers.

From our simulation of a successful CE ejection, we find that the final separation between the core of the giant star and the companion is too large to release GWs in the frequency range where LISA could detect them. This result applies to all successful CEs that we simulated, but observations seem to suggest smaller final orbits, which would move the final systems closer to the frequency range of LISA.
We also find that during successful CE ejections, the GWs emitted by the envelope contribute to the whole signal mostly in the lowest frequency range. This plays a very small role for the detection of such a system, and thus the signal associated with the envelope can safely be ignored.

Computing the GWs originating from the last stage of a CE-merger in 3D HD simulations of the interaction of the core of a giant star with a WD companion, we find their frequency inside the frequency range of LISA. They might therefore be detectable depending on the distance. We also used toy models to estimate the contribution to the GW signal from the inspiral phase and showed that it further increases the chances of detecting these systems.
Such a detection would help to better understand the CE stage. The time evolution of the GWs gives an accurate depiction of how the orbital frequency changes as the companion inspirals through the envelope. GWs could serve as probes to study the interior of giant stars, helping us model the drag forces experienced by the companion and the density profile of the star. The exact conditions required for a successful ejection of the envelope or those leading to a merger are not fully known; detections of these events would place constraints on these cases. These events would probably have a detectable electromagnetic counterpart that could make them a suitable target for multimessenger astronomy.

\section*{ACKNOWLEDGMENTS}
T.S. and J.M-F. are fellows of the International Max Planck Research School for Astronomy and Cosmic Physics at the University of Heidelberg (IMPRS-HD) and acknowledge financial support from IMPRS-HD.

This work acknowledges support by the European Research Council (ERC) under the European Union’s Horizon 2020 research and innovation programme under grant agreement No.\ 759253 and 945806, the Klaus Tschira Foundation, and the High Performance and Cloud Computing Group at the Zentrum f{\"u}r Datenverarbeitung of the University of T{\"u}bingen, the state of Baden-W{\"u}rttemberg through bwHPC and the German Research Foundation (DFG) through grant no INST 37/935-1 FUGG.

T.S. and A.B. acknowledge support by the State of Hesse within the Cluster Project ELEMENTS. A.B. also acknowledges support by Deutsche Forschungsgemeinschaft (DFG, German Research Foundation) - Project-ID 279384907 - SFB 1245.

F.R. and A.B. acknowledge support by DFG through Project-ID 138713538 - SFB 881 (``The Milky Way System'', subproject A10)
\bibliographystyle{aa}

\end{document}